\documentclass[11pt]{article}

\usepackage[margin=1in]{geometry}
\usepackage{amsmath,amssymb,amsfonts}
\usepackage{graphicx}
\usepackage{booktabs}
\usepackage{natbib}
\usepackage{hyperref}
\usepackage{xcolor}
\usepackage{enumitem}
\usepackage{bm}
\usepackage{float}
\usepackage{array}

\title{Methodological Changes to the Attention ResUNet Hourly\\
  Precipitation Postprocessor}

\author{Thomas M.\ Hamill\thanks{The Weather Company, Atlanta,
  Georgia; e-mail: \texttt{tom.hamill@weather.com}}}

\date{\today}

\begin{document}

\maketitle

\begin{abstract}
This note is a technical companion to a previously published preprint
describing an Attention Residual U-Net that postprocesses deterministic
forecasts from The Weather Company's Global and Regional Atmospheric
Forecast (GRAF) model into probabilistic hourly precipitation forecasts
\citep{hamill2026}. It documents what has changed in that method since
publication. Feature-wise Linear Modulation \citep[FiLM;][]{perez2017}
conditioning on calendar season and forecast lead time is
used to produce a single trained model for each season, replacing 192
separately trained per-month, per-lead checkpoints.  Lead time is extended
from 48 to 72\,h. Two new input channels are used, per-pixel local solar hour
and a static, monthly-varying precipitation climatology. During
verification, the climatological reference against which the Brier
Skill Score is computed now has an added diurnal dimension, on top of
the monthly resolution it already had.
Brier Skill Score and reliability are compared between
the new vs. the previous training.  Forecasts generated with the new
training show a modest, consistent improvement
of the current training over the original.
\end{abstract}

\section{Introduction and Scope}
\label{sec:intro}

This note documents what has changed in the Attention ResUNet
probabilistic precipitation postprocessor since the methodology
described in the original preprint \citep{hamill2026}. It assumes familiarity with
the domain-specific terminology (GRAF, MRMS, CONUS, etc.) established
there and does not re-define it here; acronyms introduced or
non-trivially expanded in this note itself are defined at first use
below.  While many algorithmic improvements were tested since the first 
version, this manuscript documents only the most recent configuration 
and compares that to the previously described method.

Section~\ref{sec:changes} describes what changed in the core modeling
methodology --- input features and training data, network conditioning,
training regime, and inference. Section~\ref{sec:verification}
describes what changed in how the method is verified. Both sections
are written against the original manuscript's own section structure so
they can be read side by side with it.

\section{Methodological Changes}
\label{sec:changes}

Table~\ref{tab:summary} summarizes the changes described in this
section. The core architectural ingredients of the original
manuscript --- the Attention ResUNet backbone with attention-gated skip
connections, the zero-inflated two-component Gamma mixture predictive
distribution, the negative-log-likelihood loss, the
expectation-maximization (EM)-based
climatological bias initialization, and fully-convolutional
whole-domain inference --- are all unchanged and are not repeated here.
Season labels below follow standard meteorological convention: DJF =
December/January/February, MAM = March/April/May, JJA =
June/July/August, SON = September/October/November.

\begin{table}[H]
\centering
\caption{Summary of methodological changes since the original
  manuscript. See subsections~\ref{sec:changes-features}--\ref{sec:changes-inference}
  for details.
  }
\label{tab:summary}
\medskip
\begin{tabular}{>{\raggedright\arraybackslash}p{3.3cm}>{\raggedright\arraybackslash}p{5.4cm}>{\raggedright\arraybackslash}p{5.4cm}}
\toprule
\textbf{Axis} & \textbf{Original manuscript} & \textbf{Current} \\
\midrule
Input channels &
  7: GRAF precip, terrain deviation, GFS RH, precip$\times$terrain,
  precip$\times$RH, terrain gradient (zonal, meridional) &
  10: the same 7, plus per-pixel local-solar-hour sine/cosine and a
  static precipitation climatology channel \\
\addlinespace
Lead/date conditioning &
  None; lead and calendar dependence handled implicitly by training a
  separate checkpoint for each &
  FiLM conditioning vector [sin(day-of-year), cos(day-of-year),
  lead/72h] applied at each encoder/decoder stage \\
\addlinespace
Checkpoint granularity &
  One checkpoint per (calendar month $\times$ 3-h lead step),
  or 192 checkpoints; nearest-lead weights used for
  leads that are not even multiples of 3\,h &
  One checkpoint per calendar season (DJF/MAM/JJA/SON), 4 total \\
\addlinespace
Lead-time coverage &
  +3\,h to +48\,h &
  +3\,h to +72\,h, continuously via FiLM \\
\addlinespace
Training-data pipeline &
  One precomputed patch set per (month, lead) checkpoint, drawn from
  four 60-day recency/seasonal date windows &
  Precomputed, season-and-lead-pooled zarr patch archives spanning all
  years and all leads within a season, read with a chunk-aware
  shuffle sampler \\
\addlinespace
Inference procedure &
  Fully-convolutional single forward pass over the padded CONUS domain
  &
  Same, but only achievable for the FiLM-conditioned model because
  solar-hour is a per-pixel input rather than a spatially-varying FiLM
  term (\S\ref{sec:changes-inference}) \\
\bottomrule
\end{tabular}
\end{table}

\subsection{Input Features and Training Data}
\label{sec:changes-features}

The original manuscript's seven input channels (its Section 3.1,
``Input Features and Training Data'') are all retained unchanged. Two
channels have been added:

\begin{enumerate}[label=(\arabic*)]
  \item \textbf{Local solar hour} (sine and cosine): the valid-time
        local solar hour at each pixel, computed from the UTC valid
        hour and the pixel's own longitude ($\text{solar\_hour} =
        (\text{UTC hour} + \lambda/15) \bmod 24$, then sine/cosine
        encoded). This is intended to capture a possible diurnal 
        dependence of convective precipitation errors, which the 
        original manuscript's feature set did not represent at all.
  \item \textbf{Precipitation climatology}: a static, per-pixel,
        month-indexed monthly precipitation climatology (a blend of
        PRISM \citep[Parameter-elevation Regressions on Independent
        Slopes Model;][]{daly2008}, WorldClim \citep{fick2017}, and
        ERA5 \citep[the ECMWF fifth-generation atmospheric
        reanalysis;][]{hersbach2020}, following the same recipe used
        for a companion High-Resolution Rapid Refresh
        \citep[HRRR;][]{benjamin2016, dowell2022} postprocessing
        project), log-transformed and
        normalized to $[0,1]$. This gives the network an explicit prior
        on which locations and months are climatologically wetter or
        drier, independent of the current forecast.
\end{enumerate}

The use of both channels is motivated by the same design principle already used
in the original manuscript for the terrain-interaction and
humidity-interaction channels: rather than relying on the network to
learn a relationship implicitly from many training examples, a
physically motivated derived quantity is computed explicitly and
handed to the network as an input.

The training-data pipeline itself also changed. Both the original and
current schemes precompute their training patches offline rather than
sampling on the fly during training; what changed is the granularity
of that pre-computation. The original manuscript's data-sampling scheme
(its Section 3.1.4) precomputes and saves one patch set per (month,
lead) checkpoint --- up to 192 of them --- using the macro/micro
wet-day-weighted sampling described below, drawn from four 60-day
recency/seasonal date windows; training simply loads and cycles
through that pre-saved set. The current pipeline instead precomputes
zarr-backed patch archives, one per calendar season, pooling patches
from every
available year of that season and every lead time (rather than one
(month, lead)-specific set per checkpoint). This is what makes the 
seasonally pooled, FiLM-conditioned training scheme in
\S\ref{sec:changes-conditioning}--\ref{sec:changes-training} possible;
a single season's training set now needs to contain patches spanning
the full 3--72\,h lead range, not just one lead.  Some advantages of this
approach include: (a) fewer training weights to manage, and (b) a larger training
sample size through compositing training across lead times, resulting
in presumably more stable, higher-quality weights.  The macro/micro
wet-day-weighted patch selection described in the original manuscript
(more samples on wet days, importance sampling toward higher
precipitation) is preserved at patch-pool-write time. Training/holdout
partitioning is now made explicit as a 7-day-block split (contiguous
one-week blocks assigned entirely to training or holdout, so that nearby
days within the same synoptic episode cannot leak across the split).

\subsection{Network Conditioning (FiLM)}
\label{sec:changes-conditioning}

Feature-wise Linear Modulation (FiLM) is the most consequential 
change that was added to the Attention ResUNet.
A small conditioning MLP maps a low-dimensional conditioning vector to
a per-channel scale and shift (gamma, beta) applied after each of the
encoder, bridge, and decoder stages. The conditioning vector is three
scalars, constant over the whole domain for a given forecast:
\[
  \text{cond} = \big[\,\sin(\text{doy}),\ \cos(\text{doy}),\
  \text{lead}/72\text{h}\,\big]
\]
where $\text{doy}$ is the forecast's calendar day of year (1--366 for
the initialization date), sine/cosine encoded so the conditioning
signal varies smoothly across the December/January boundary.
This lets one model, trained on patches pooled across every lead time
within a season, still express lead- and calendar-time-dependent
behavior.

In the new version, the solar hour was promoted out of the conditioning vector
entirely and into the two new per-pixel input channels described in
\S\ref{sec:changes-features} above.   This kept the ability
to achieve single-pass, whole-domain inference.

\subsection{Training Regime}
\label{sec:changes-training}

The original manuscript trains one checkpoint per calendar month per
3-hourly lead step from +3 to +48\,h, or 192
checkpoints, warm-starting each lead's training from the nearest
already-trained lead. The current scheme instead trains one checkpoint
per calendar season (DJF, MAM, JJA, SON), or 4 checkpoints total.
Each covers the full +3 to +72\,h lead range continuously via the
FiLM lead term. This is a direct consequence of the season-pooled
training data and FiLM conditioning described above: because a single
training set now spans all leads within a season, a single model can
be trained on it, rather than needing one model per lead.

The core optimization hyperparameters are unchanged: Adam with the same
base learning rate, the same \texttt{ReduceLROnPlateau} schedule, the
same early-stopping patience, the same gradient-clipping norms, and the
same negative-log-likelihood loss (Gamma-mixture NLL) with the same
numerical-stability safeguards described in the original manuscript's
Section 3.5.2. The season trainer imports these settings
directly from the original per-lead trainer rather than re-deriving
them, so this is a change in \emph{what} is pooled for training, not in
\emph{how} the optimization itself is configured.

\subsection{Inference Procedure}
\label{sec:changes-inference}

Both the original and current methods perform inference as a single
forward pass over the whole (edge-replication-padded) CONUS domain,
exploiting the fact that the Attention ResUNet is fully convolutional.
This property itself has not changed and is not new. What changed is
how the correct checkpoint and conditioning are selected for a given
forecast: the original manuscript selects among up to 192 checkpoints
by nearest 3-hourly lead to the requested date and lead; the current
method selects one of 4 season checkpoints by calendar month and
supplies the requested lead (and day-of-year) directly to the FiLM
conditioning vector, with no nearest-neighbor approximation needed
across lead time.

\section{Verification and Evaluation Methodology Changes}
\label{sec:verification}

The verification metrics themselves --- reliability diagrams, Brier
Skill Score (BSS) against a climatological reference, and performance
diagrams, all as described in the original manuscript's Section 6 ---
are unchanged, as is the terrain-roughness stratification of the results
 (top-10\% roughest terrain vs.\ the remaining 90\%) used
in its Results section. Two things have changed: the climatological
reference standard used to compute BSS (\S\ref{sec:verification-climo}),
and the evaluation infrastructure used to compare successive versions
of the trained model against each other, which did not exist at all
when the original manuscript was written
(\S\ref{sec:verification-crosscheck}).

\subsection{Climatological Reference Refinement}
\label{sec:verification-climo}

The original manuscript computes its BSS reference from NCEP Stage IV
2020--2024 hourly precipitation analyses, bilinearly interpolated to
the GRAF grid (its Section 6), with no explicit
diurnal (hour-of-day) resolution, only a monthly one. Two things about
this have since changed. First, the interpolation itself was redone
with a Delaunay triangulation and linear barycentric interpolation
rather than simple bilinear interpolation.  Second, the climatology gained
an explicit diurnal dimension (all 24 UTC hours, not just a monthly
mean), since GRAF gamma-mixture inference runs at 1--3\,h lead
steps and a monthly-only climatology cannot resolve the diurnal cycle
of convection.

\subsection{Comparing Forecast Skill: Original vs.\ Current Training}
\label{sec:verification-crosscheck}

The original manuscript verifies its model against a single reference
forecast (Gaussian-smoothed, thresholded GRAF).  Here, a 
comparison against this reference standard is omitted, and we compare
only the forecasts from the newly trained version and the older trained version.
Both models are scored on the same Mar/Jun/Sep/Dec 2025,
6-hourly sample used for the original manuscript's own BSS and
reliability figures (447 initialization times per lead with complete
GRAF/MRMS/probability data).

\begin{figure}[H]
\centering
\includegraphics[width=0.85\textwidth]{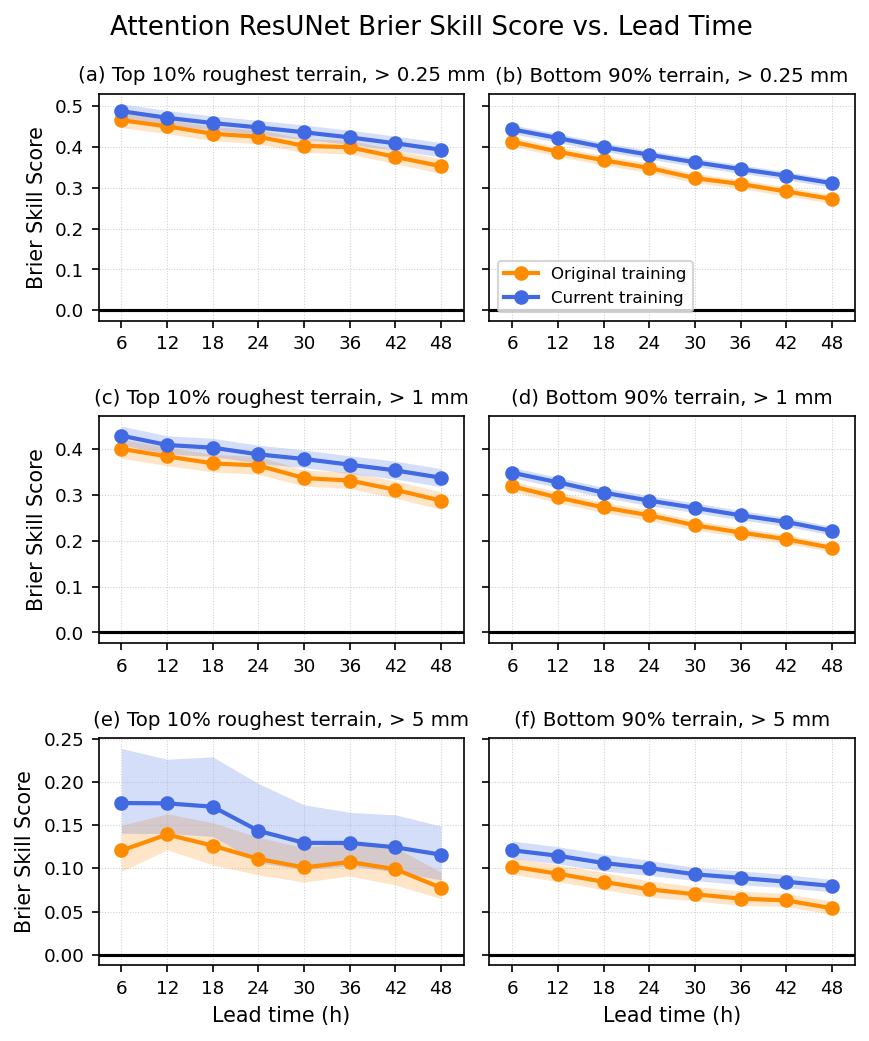}
\caption{Brier Skill Score vs.\ lead time for the Attention ResUNet's
  own postprocessed (Gamma-mixture) probabilities, original training
  (orange) vs.\ current training (blue), stratified by terrain
  roughness as in the original manuscript's own BSS figure. Shaded
  bands are 90\% Hodges-Lehmann confidence intervals (see text).
  Sample: 2025 March/June/September/December, 6-hourly, matching the
  original manuscript's own verification sample.}
\label{fig:bss_compare}
\end{figure}

Each curve in Figure~\ref{fig:bss_compare} is shown with a shaded 90\%
confidence interval on that model's own per-cycle BSS distribution
(one value per 6-hourly verification cycle, pooling all grid points
within that cycle into a single score, following the case-pooling
philosophy of \citet{hamill1999}). The interval is the Hodges-Lehmann
nonparametric confidence interval obtained by inverting the
(one-sample) Wilcoxon signed-rank statistic on those per-cycle values.
Because that statistic targets the median of the per-cycle BSS \emph{ratios},
which differs from the pooled (ratio-of-sums) BSS actually plotted, the band
is recentered on the plotted score: its rank-based half-widths are
kept, but shifted so the shaded region is centered on the curve it
shades rather than on the Hodges-Lehmann location estimate itself.

Figure~\ref{fig:bss_compare} shows the terrain-roughness-stratified
BSS-vs.-lead-time comparison. The current training scores at or above
the original training at every lead time, threshold, and terrain
stratum shown, with the largest relative gains at 5\,mm over the
roughest 10\% of terrain --- e.g., at 6\,h, BSS improves from 0.121 to
0.176, and at 48\,h from 0.078 to 0.116 --- precisely the
heavy-precipitation-over-complex-western-terrain regime the original
manuscript highlights as the hardest case. Both curves and their CI
bands are well separated and well behaved at every threshold in both
terrain strata. Gains at 0.25\,mm and
1\,mm are more modest but still consistent, e.g.\ at 48\,h and
0.25\,mm, BSS improves from 0.353 to 0.393 over the roughest terrain
and from 0.272 to 0.311 over smoother terrain.

\begin{figure}[H]
\centering
\includegraphics[width=\textwidth]{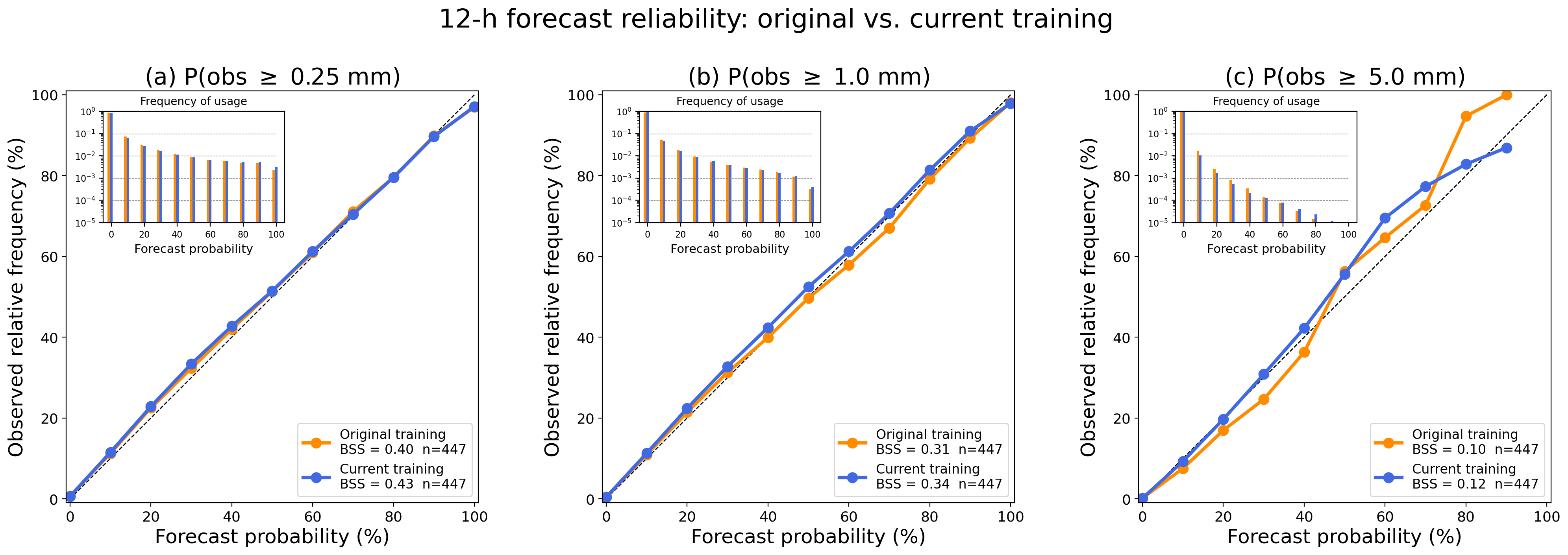}
\caption{Reliability diagram for the Attention ResUNet's own
  postprocessed (Gamma-mixture) probabilities, original training
  (orange) vs.\ current training (blue), at +12\,h lead, for
  exceedance of 0.25, 1.0, and 5.0\,mm. Inset panels show
  forecast-probability usage frequency. Sample as in
  Figure~\ref{fig:bss_compare}.}
\label{fig:relia_compare_12h}
\end{figure}

\begin{figure}[H]
\centering
\includegraphics[width=\textwidth]{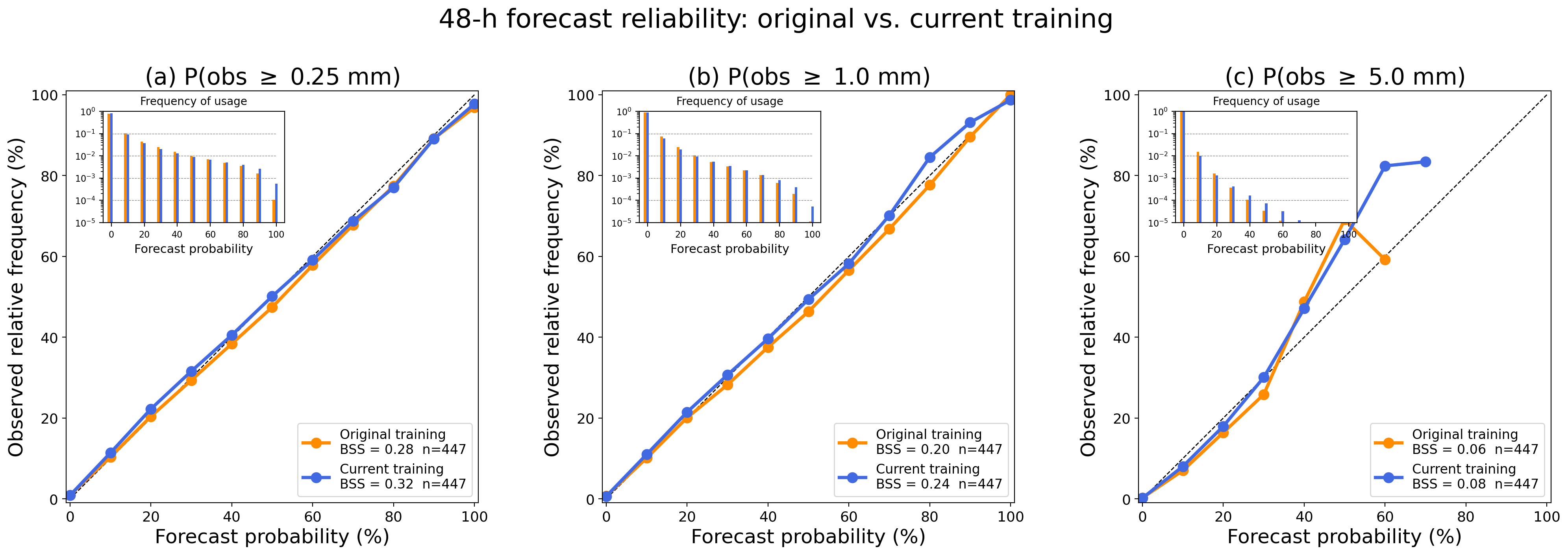}
\caption{As Figure~\ref{fig:relia_compare_12h}, but at +48\,h lead.}
\label{fig:relia_compare_48h}
\end{figure}

Figures~\ref{fig:relia_compare_12h} and~\ref{fig:relia_compare_48h}
show the corresponding (unstratified, CONUS-wide) reliability diagrams
at +12\,h and +48\,h --- the same two lead times shown in the original
manuscript's own reliability figures. Both trainings are well calibrated at
both lead times; the current training's BSS is above the original
training's at every threshold shown: at +12\,h, 0.43 vs.\ 0.40 at
0.25\,mm, 0.34 vs.\ 0.31 at 1\,mm, and 0.12 vs.\ 0.10 at 5\,mm; at
+48\,h, 0.32 vs.\ 0.28 at 0.25\,mm, 0.24 vs.\ 0.20 at 1\,mm, and 0.08
vs.\ 0.06 at 5\,mm. This is consistent with the lead-time comparison
in Figure~\ref{fig:bss_compare} and with a modest, genuine improvement
in the postprocessed forecast rather than sampling noise, though the
5\,mm curves at both lead times still show some sampling irregularity
at the highest forecast probabilities, where usage frequency (inset
panels) drops below $10^{-3}$.

\begin{figure}[H]
\centering
\includegraphics[width=\textwidth]{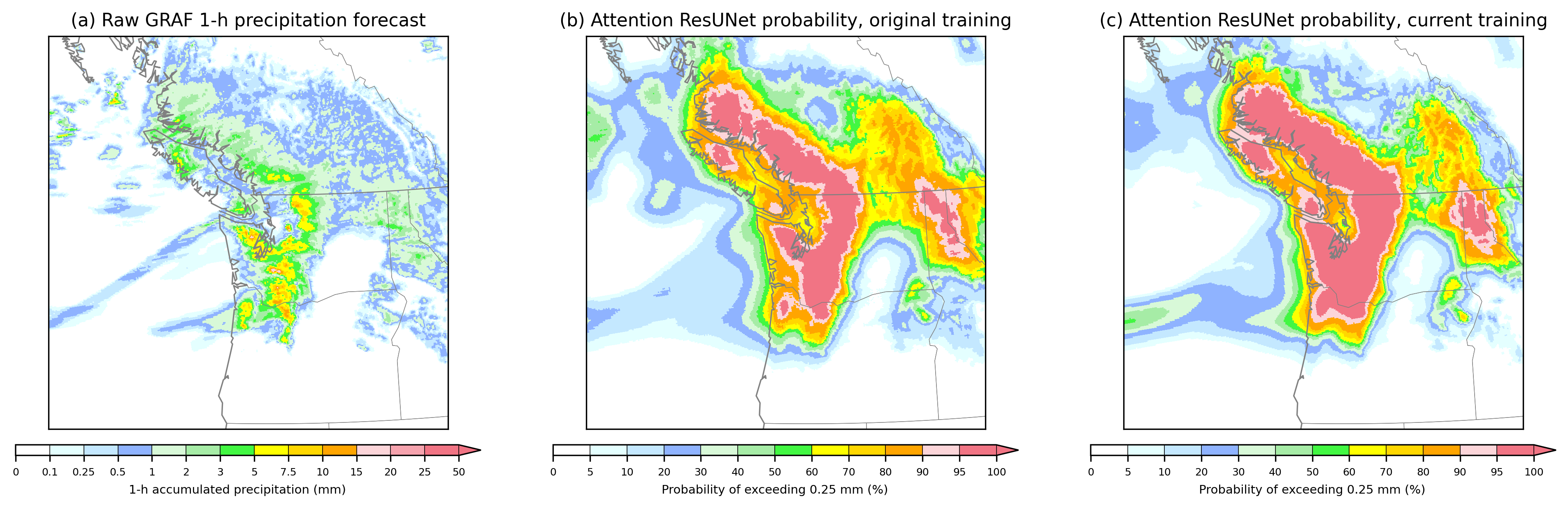}
\caption{Washington-state atmospheric river case, IC 2026031112, +12\,h.
  (a) Raw GRAF 1-h accumulated precipitation forecast, (b) Attention
  ResUNet probability of exceeding 0.25\,mm, original training, and
  (c) the same, current training.}
\label{fig:case_day}
\end{figure}

Figure~\ref{fig:case_day} recreates, for the current comparison, one
of the original manuscript's own case-study figures: its Washington
State atmospheric river case. The
case day shown here is a separate event from the one in the original
manuscript (2026-03-12 00Z, +12\,h, selected as the strongest 2026
box-mean 1-h MRMS accumulation over the Olympic Peninsula/Cascades
region), since both models needed to have inference available for the
exact same case for the comparison to be meaningful, and the original
manuscript's own case day predates the current training. Both
trainings reproduce the original manuscript's qualitative finding that
the postprocessed probability field is linked to terrain rather than
simply following the raw GRAF forecast (panel a): high probabilities
hug the Olympic Peninsula and Cascade Range in both panel (b) and
panel (c), unlike the offshore-centered maximum in the raw forecast.
The clearest difference between the two panels is the extension of areal
coverage of high probability.

\section{Summary}
\label{sec:summary}

This note describes recent updates to a deep learning precipitation postprocessing
method, previously documented in \citep{hamill2026}.
The core statistical method described in the original manuscript is unchanged.
It is an Attention ResUNet predicting a zero-inflated two-component Gamma
mixture at each pixel, trained by negative log-likelihood with
climatological bias initialization, run as a single fully-convolutional
forward pass over the CONUS domain. What has changed
since that manuscript was written is how the model is conditioned and
trained across lead times and calendar seasons, and, separately, how
verification of that model is done.

On the modeling side, FiLM conditioning now lets one model per calendar
season cover a continuous +3 to +72\,h lead range, replacing up to 192
separately trained per-(month, lead) checkpoints covering only +3 to
+48\,h; this directly implements a specific improvement the original
manuscript's own Summary section proposed as future work. Two input
channels were added (per-pixel local solar hour, and a static
precipitation climatology), and the training-data pipeline moved from
per-checkpoint on-the-fly sampling to precomputed, season-and-lead-pooled
patch archives. Single-pass whole-domain inference, a feature already
present in the original manuscript, was preserved through this change
only because a spatially-varying conditioning term (solar hour) was
demoted from the FiLM vector to a per-pixel input channel; had that not
been done, the FiLM-conditioned model would have required a return to
patch-tiled inference.

On the verification side, the climatological reference used for Brier
Skill Score was consolidated into one canonical, diurnally-resolved
file shared with a companion HRRR project (a provenance and resolution
improvement, not a large numerical correction), and a new
cross-checkpoint reliability-comparison capability was built so that
successive retrains of the model can be compared against each other
using cached scores, something the original single-model-version
manuscript had no occasion to do.

Newly trained forecasts were shown to retain the general characteristic of reliability,
while forecast skill with respect to a climatological reference was improved.


\end{document}